\documentclass[twocolumn]{aastex62}
\newcommand{\vdag}{(v)^\dagger}

\usepackage{soul}
\newcommand\Swift{\emph{Swift} }
\newcommand\Fermi{\emph{Fermi} }
\graphicspath{{./}{figures/}}
\shorttitle{\Fermi Unassoc with Swift-XRT}
\shortauthors{Kaur et al.}
\begin{document}
\title{Classification of New X-ray Counterparts for \Fermi
Unassociated Gamma Ray Sources Using the \Swift X-Ray Telescope}

\author[0000-0002-0878-1193]{Amanpreet Kaur}
\affil{The Pennsylvania State University, 525 Davey Lab, University Park, PA 16802, USA}

\author{Abraham D. Falcone}
\affil{The Pennsylvania State University, 525 Davey Lab, University Park, PA 16802, USA}

\author{Michael D. Stroh}
\affil{Center for Interdisciplinary Exploration and Research in Astrophysics (CIERA), Northwestern University, Evanston, IL 60208, USA}

\author{Jamie A. Kennea}
\affil{The Pennsylvania State University, 525 Davey Lab, University Park, PA 16802, USA}

\author{Elizabeth C. Ferrara}
\affil{NASA Goddard Space Flight Center, Greenbelt, MD 20771, USA}
\affil{Department of Astronomy, University of Maryland College Park, MD 20742, USA}

\begin{abstract}
Approximately one-third of the gamma-ray sources in the third \Fermi-LAT catalog are unidentified or unassociated with objects at other wavelengths. Observations with the X-Ray Telescope on the Neil Gehrels \Swift Observatory (\Swift-XRT) have yielded possible counterparts in $\sim$30\% of these source regions. The objective of this work is to identify the nature of these possible counterparts, utilizing their gamma ray properties coupled with the \Swift derived X-ray properties. The majority of the known sources in the \Fermi catalogs are blazars, which constitute the bulk of the extragalactic gamma-ray source population. The galactic population on the other hand is dominated by pulsars. Overall, these two categories constitute the majority of all gamma-ray objects. Blazars and pulsars occupy different parameter space when X-ray fluxes are compared with various gamma-ray properties. In this work, we utilize the X-ray observations performed with the \Swift-XRT for the unknown \Fermi sources and compare their X-ray and gamma-ray properties to differentiate between the two source classes. We employ two machine learning algorithms, decision tree and random forest classifier, to our high signal-to-noise ratio sample of 217 sources, each of which correspond to Fermi unassociated regions. The accuracy score for both methods were found to be 97\% and 99\%, respectively. 
The random forest classifier, which is based on the application of a multitude of decision trees, associated a probability value (P$_{bzr}$) for each source to be a blazar. This yielded 173 blazar candidates from this source sample, with P$_{bzr}$ $\geq$ 90\% for each of these sources, and 134 of these possible blazar source associations had P$_{bzr}$ $\geq$ 99\%. The results yielded 13 sources with P$_{bzr}$ $\leq$ 10\%, which we deemed as reasonable candidates for pulsars, 7 of which result with P$_{bzr}$ $\leq$ 1\%. There were 31 sources that exhibited intermediate probabilities and were termed ambiguous due to their unclear characterization as a pulsar or a blazar.

\end{abstract}


\section{Introduction} \label{sec:intro}
Since the launch of the \Fermi Gamma Ray Space Telescope in June 2008, thousands of gamma-ray sources have been discovered in our universe. Four point source catalogs have been published to-date, with 1451 sources in the 1FGL\citep{Abdo2010a} catalog, 1873 sources in 2FGL \citep{Nolan2012} catalog, and 3033 sources in the 3FGL\citep{Acero2015} catalog; as well as 5065 sources in the recently released 4FGL, which is too recent to be considered in the multi-wavelength follow-up and classification effort that is described in this paper. The dominant source classes in all of these catalogs are blazars and pulsars, representing the extragalactic and galactic sky, respectively. Other classes include X-ray binaries, gamma ray bursts, supernova remnants, globular clusters, starburst galaxies, etc. Most of the sources in the 1FGL and 2FGL catalogs are also present in the 3FGL catalog, with much improved measurements ($\sim$ 2.5' uncertainty). While some of these sources are attributed to one or the other class, about one-third (1010) are unassociated and unidentified. A rather large fraction of the known gamma-ray sources are blazars (75\%), therefore it is highly likely that some of the unassociated ones could belong to a fainter subclass of blazars. Finding these blazars would offer an opportunity to conduct the population studies in a complete manner, thereby shedding light on the still debated idea of a blazar sequence \citep{Fossati1998,Ghisellini2017}. In addition to blazars, some previous studies of unassociated sources from \Fermi catalogs have led to discoveries of millisecond pulsars, black widows, redback pulsars, high mass X-ray binaries, and extreme blazars; e.g., See \citet{Parkinson2010,Ransom2010}. The emission processes of these newly discovered objects are still not completely understood and are an active field of research. Furthermore, some of these objects could potentially be the candidates for a new class of gamma-ray sources, which could help to uncover new and extreme astrophysical environments that could possibly contribute to studies of new physics. Overall, finding the nature of these mysterious gamma-ray sources is critical for furthering our understanding of gamma-ray blazar and puslar systems, as well as possible new source classes, and for the study of the gamma-ray sky and the extreme environments that illuminate it. Finding and classifying multiwavelength counterpart sources is a logical first step in this process. \\
In the past, \citep{Massaro2012} developed a technique, further refined by \citep{DAbrusco2013} which utilized WISE \citep{Wright2010} colors to differentiate blazars from other source populations. However, to identify both pulsars and blazars, various machine learning algorithms were  successfully employed utilizing the Fermi-LAT gamma-ray data, e.g., see  \citet{Parkinson2016}, \citep{Lefaucheur2017}. In this work, we attempt to characterize the new potential associations for the 3FGL unassociated sources that have been found by \citet{Falcone2019} by applying machine learning algorithms to their X-ray and gamma-ray parameters obtained from Fermi and \Swift-XRT observations of these regions, respectively. The reason for utilizing X-ray observations is based on the fact that the gamma-ray and X-ray bands are close enough in energy space to share many of the same types of high energy emitters as their source populations. Moreover, the X-ray observations with \Swift reduces the positional uncertainty of these \Fermi sources from a few arcminutes to a few arcseconds, thereby making the identification process much easier. More importantly, pulsars and blazars occupy different parameter space when X-ray fluxes are compared \citep{Falcone2015}, which makes it a crucial parameter for machine learning algorithms to classify sources as blazars or pulsars. The structure of this paper is described as follows: Section~\ref{sec:sample} describes the observational details and sample selection criteria. In addition, the details of analysis procedure are explained in this section. Section~\ref{sec:methods} describes our findings by comparing gamma-ray and X-ray properties of our sample. In Section~\ref{sec:ml}, we introduce machine learning methods employing gamma-rays and X-rays to classify these objects as blazars or pulsars. A detailed discussion of our conclusions are described in Section~\ref{sec:conclusion}.

\section{Observations and Analysis}\label{sec:sample}
A sample of unidentified objects from the 3FGL catalog were selected for observations with \Swift-XRT through \Swift fill-in and GI programs to find potential X-ray counterparts. Detailed information about the sample selection, observations, and analysis methods can be found in \citet{Falcone2019}. One of the selection criteria for this sample was based on the desire to contain the confidence regions of the 3FGL sources within the field-of-view of \Swift-XRT. Therefore, the sources with position confidence region semi-major axis $<$ 10' were selected.  At the time of this writing, the total sample included 803 targeted 3FGL positions. The exposure time for each source was typically $\sim$ 4\,ksec. \\
From the 803 unassociated \Fermi sources that were observed, at least one X-ray source was detected in 552 of the the 95\% uncertainty regions. For this study, the following two selection criterian were utilized: (i) only the objects with detections at the significance threshold of Signal-to-Noise ratio $\geq$ 4, and (ii) the sources with only one X-ray counterpart within the 95\% Fermi confidence region were selected. This led to a total of 217 X-ray sources found within the 95\% confidence regions of 217 \Fermi unassociated sources. The complete details of these 217  sources are provided in \citet{Falcone2019}. \\

\section{Methods}\label{sec:methods}
\begin{figure*}
\centering
\includegraphics[trim=1.0cm 0.0cm 0.0cm 0.0cm,angle=0,width=\textwidth]{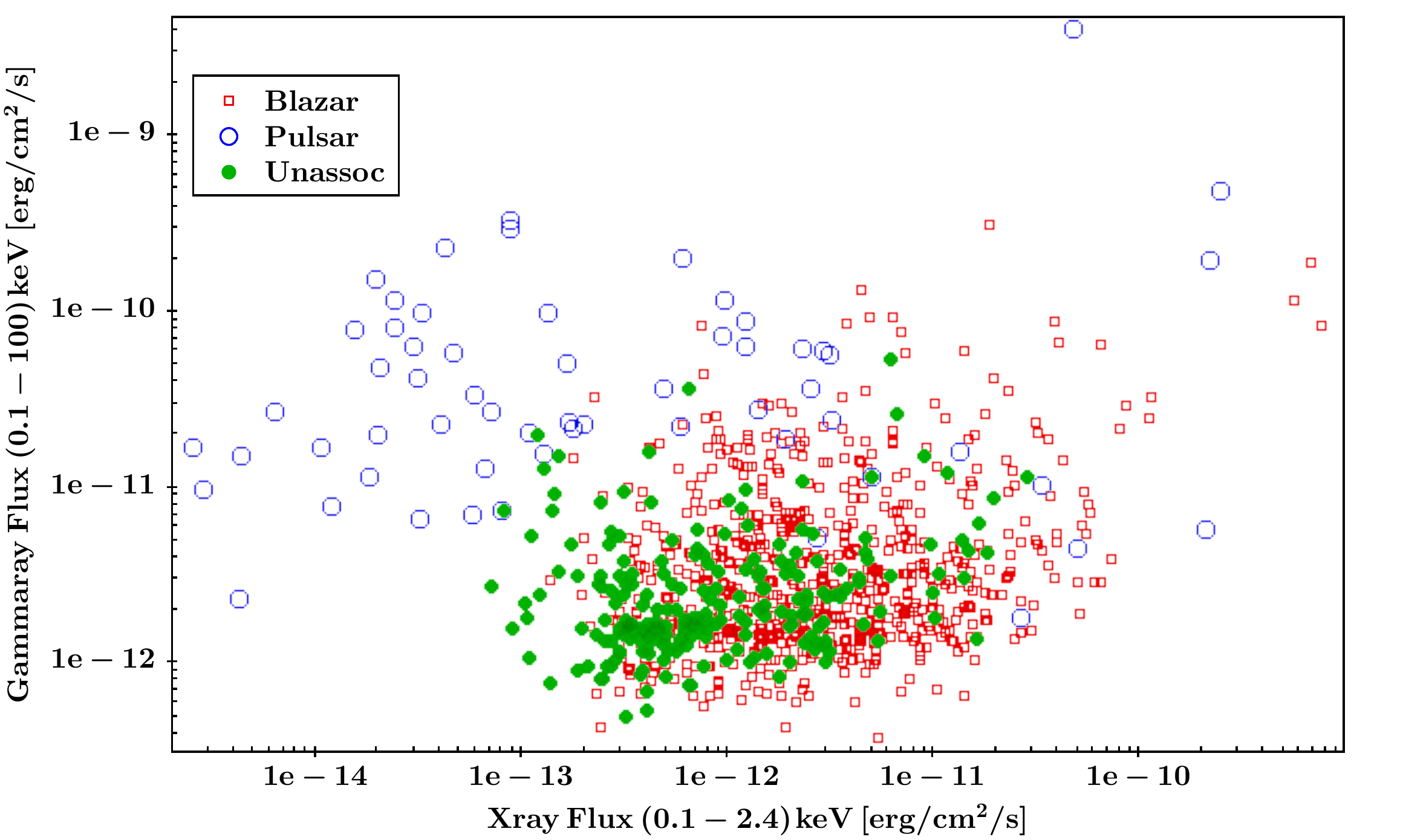}
\caption{X-ray vs gamma-ray flux from known blazars ($red$) and pulsars ($blue$). The 217 unassociated sources ($green$) are plotted over the same space \label{fig:GammaXray}}
\end{figure*}

\begin{figure*}
\centering
\includegraphics[trim=1.0cm 0.0cm 0.0cm 0.0cm,angle=0,width=\textwidth]{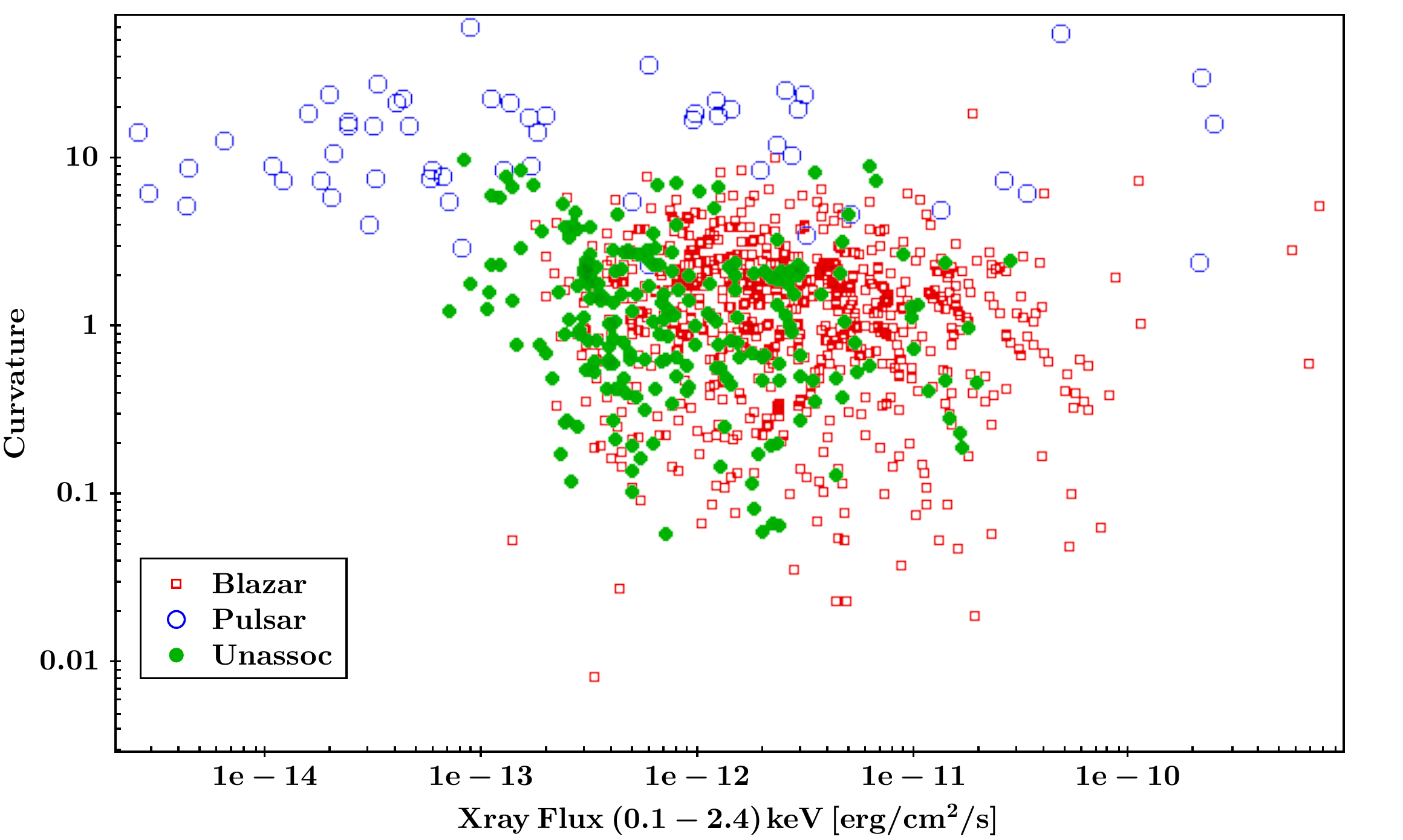}
\caption{X-ray flux vs curvature index from known blazars ($red$) and pulsars ($blue$). The 217  unassociated sources ($green$) are plotted over the same space\label{fig:curId}}
\end{figure*}

\begin{figure*}
\centering
\includegraphics[trim=1.0cm 0.0cm 0.0cm 0.0cm,angle=0,width=\textwidth]{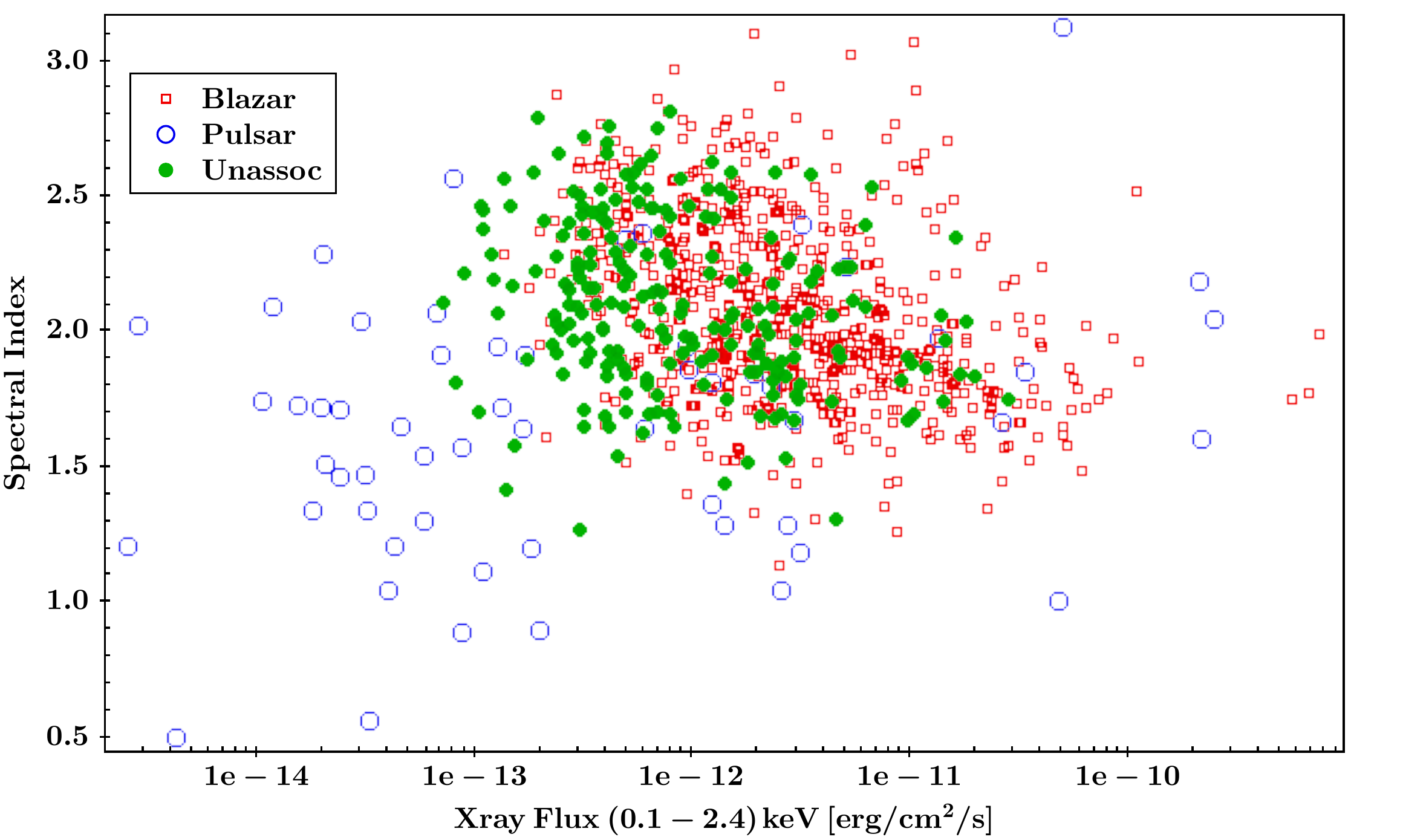}
\caption{X-ray flux vs spectral index from known blazars ($red$) and pulsars ($blue$). The 217 unassociated sources ($green$) are plotted over the same space\label{fig:spId}}
\end{figure*}

\begin{figure*}
\centering
\includegraphics[trim=1.0cm 0.0cm 0.0cm 0.0cm,angle=0,width=\textwidth]{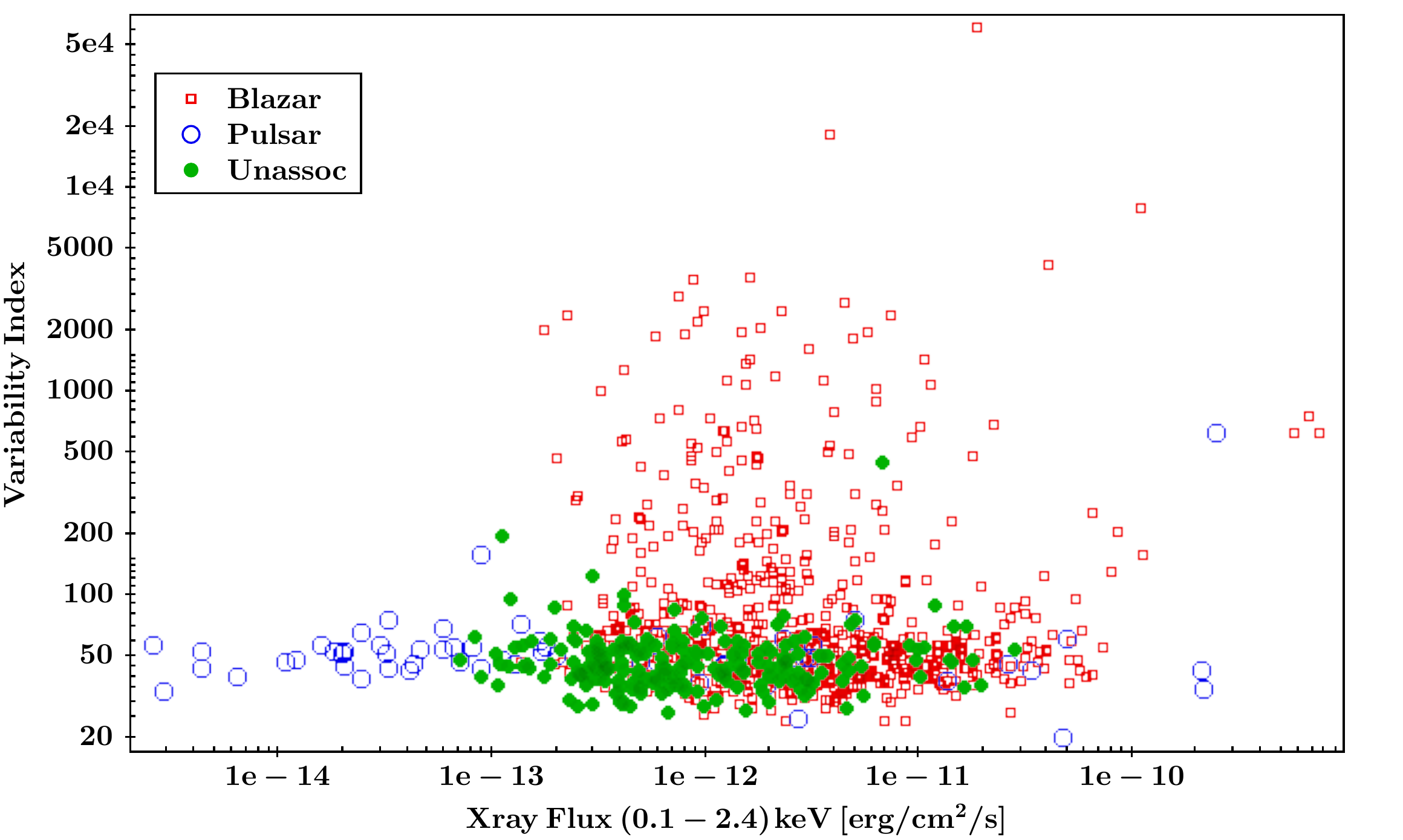}
\caption{X-ray flux vs variability index from known blazars ($red$) and pulsars ($blue$). The 217 unassociated sources ($green$) are plotted over the same space\label{fig:varId}}
\end{figure*}

The 3FGL catalog is comprised of blazars, pulsars, supernova remnants, starburst galaxies, gamma ray bursts, globular clusters etc., among the known classes of astrophysical sources. However, blazars and pulsars dominate the extragalactic and galactic source class populations, constituting $\sim$ 75\% and $\sim$ 8\% of the total sources, respectively. Therefore, it is highly likely that a majority of the unknown sources are potentially blazars or pulsars. \citet{Falcone2015} demonstrated that blazars and pulsars occupy different parameter space when gamma-ray properties are compared with X-ray fluxes. We investigate this scenario by comparing the gamma-ray and X-ray properties of the unassociated sources with that of the known blazars and pulsars.\\
The first step was to conduct a search for blazars and pulsars in literature for which both gamma-ray and X-ray data were available. Gamma-ray properties for all the known sources, i.e. known blazars and pulsars were derived from the 3FGL catalog. The X-ray flux values for blazars were acquired from the 3LAC catalog \citep{Ackermann2015b}, whereas for pulsars, X-ray fluxes were obtained from  \citet{Marelli2012},  \citet[and references therein]{Pryal2015}, \citet{ Parkinson2016, Wu2018, Zyuzin2018} and \Swift-XRT archive (See appendix for details on this analysis) . This resulted in a sample size of 753 sources; 691 blazars and 59 pulsars for which both gamma-ray data as well as typical X-ray flux were available. The number of pulsars we found in literature for which gamma-ray and X-ray observations were present relevant to this work were rather small in number as compared to blazars. 38 of these pulsars are young, 4 are middle aged and 17 are milli-second pulsars. For 217 sources in the unassociated sample, the \Swift-XRT count rate was converted to X-ray flux assuming an absorbed powerlaw spectrum with spectral index 2.0 employing PIMMS\footnote{ https://heasarc.gsfc.nasa.gov/docs/software/tools/pimms.html} tool \citep{Mukai1993}. For each source, the neutral hydrogen column density was calculated using the \href{https://heasarc.gsfc.nasa.gov/cgi-bin/Tools/w3nh/w3nh.pl}{HEASARC N$_{H}$ calculator}. \\
 The typical X-ray fluxes for pulsars are about 10-10000 times lower than gamma-ray fluxes \citep{Marelli2011}, which provides the preliminary discrimination for blazars and pulsars, as shown in Fig.~\ref{fig:GammaXray}. Moreover, the overall shape of spectral energy distribution of pulsars are more curved than blazars, which provides yet another factor for this difference, e.g., see Fig~\ref{fig:curId}. This separation can also been seen when one compares other gamma-ray properties, such as spectral indices and variability indices, as demonstrated in Fig.~\ref{fig:spId} and Fig.~\ref{fig:varId}, respectively. \\
While a comparison between gamma-ray and X-ray properties of blazars and pulsars does allow one to distinguish blazars from pulsars in a two parameter space environment, a more robust analysis is desired in order to combine all these parameters and utilize them simultaneously for the discrimination between the two dominant classes. For this purpose, we applied two machine learning classifiers as described below in Section.~\ref{sec:ml}.

\subsection{Classification with Machine Learning }\label{sec:ml}
In the last decade, although the number of gamma-ray sources have increased by a substantial amount, the number of sources with no classification has also increased. One of the best approaches to classify these objects is to obtain multi-wavelength data to create complete spectral energy distributions and thereby studying their properties in a detailed manner. This kind of work requires multiple years of investigation, thereby making it inefficient with respect to time. Recently, the big data revolution in astrophysics has motivated the community to start applying machine learning techniques for classification purposes, e.g., \citet{Ackermann2012,Mirabal2012,Mirabal2016,Parkinson2016,Salvetti2017} applied various machine learning classifiers in the context of \Fermi unidentified sources. Among all the methods employed by these authors, Random Forest Classifier \citep{Breiman2001} yielded results with accuracy $>$95\%.  We, therefore utilize a random forest classifier technique for the classification purpose in this work. For comparison and verification of the random forest results, we employed another method called Decision Tree (DT) \citep{Quinlan1990}, which is based on the same principle as the former method. A brief explanation of both methods is provided below: 
\subsubsection{Decision Tree}
A decision tree classifier (DT) is an example of a non-parametric  supervised machine learning method. It utilizes multiple given parameters to distinguish between classes by branching these parameters, one at a time, into different nodes and thereby labeling a source to one or the other class. This decision of branching/splitting is based on an index called the Gini impurity index. This index represents the probability for a source to  be assigned a wrong label/class, given it is chosen randomly from the given dataset. The nodes in the decision tree are split until a Gini impurity reaches its minimum, and at this stage, a source is labeled with the correct class. This algorithm was employed through \texttt{sklearn 0.20.3} which is available in Python3.7.3. 
\subsubsection{Random Forest}
The Random Forest (RF) method is the most commonly employed supervised technique for classification purposes. The underlying principle for RF is the decision tree method described above. The main difference in this case is that RF employs a multitude of decision trees instead of relying on the results of one such tree. The final source class is defined by taking an aggregate of the results from all these decision trees. Since, this method is based on taking an average of multiple decision tree algorithms, it provides a more robust analysis and also solves the problem of overfitting, which is commonly seen in Decision Tree methods. We used this method using \texttt{sklearn 0.20.3} which is available in Python3.7.3. utilizing 1000 decision trees and Gini inquality as the criteria for splitting the nodes for classification. The minimum number of nodes were set to 1. \\
The application of these two methods and their results are discussed below.
\subsection{Training and Test Samples}
First, the total sample (774 sources) of known blazars and pulsars for which we have \Fermi and X-ray data were divided into training and test samples; the combined training plus test sample contained 710 blazars and 64 pulsars with known characteristics. The training dataset contained 669 sources; 620 blazars and 49 pulsars. The rest of the 100 sources (90 blazars and 10 pulsars) were assigned to the test sample. The purpose of dividing the known sources into two samples is to check the accuracy of each method through the test sample after the classifier is trained on the training sample. The five parameters chosen for classification purposes were gamma-ray flux, X-ray flux, gamma ray spectral index, gamma ray variability index and curvature. These properties have already shown promise for distinguishing blazars from pulsars, as explained in Section~\ref{sec:methods}. Since the training sample is obviously biased towards one class (blazars), we employed a method called SMOTE (Synthetic Minority Over-sampling Technique) \citep{Chawla2002}, which generates synthetic data points for the under-represented class using k-nearest neighbors algorithm, choosing six as the number of nearest neighbors. We employed this algorithm utilizing Python 3.7.3. After employing this method, the training set constituted 620 blazars and 620 pulsars. In the next step, both the decision tree and random forest classifiers were run on this training set, independently. The trainer classifiers in each case were then applied to the test sample, which yielded an accuracy of 97\% and 99\% in the DT and RF cases, respectively.\\

\section{Classification Results}\label{sec:results}
\begin{figure*}
\centering
\includegraphics[trim=1.0cm 0.0cm 0.0cm 0.0cm,angle=0,width=\textwidth]{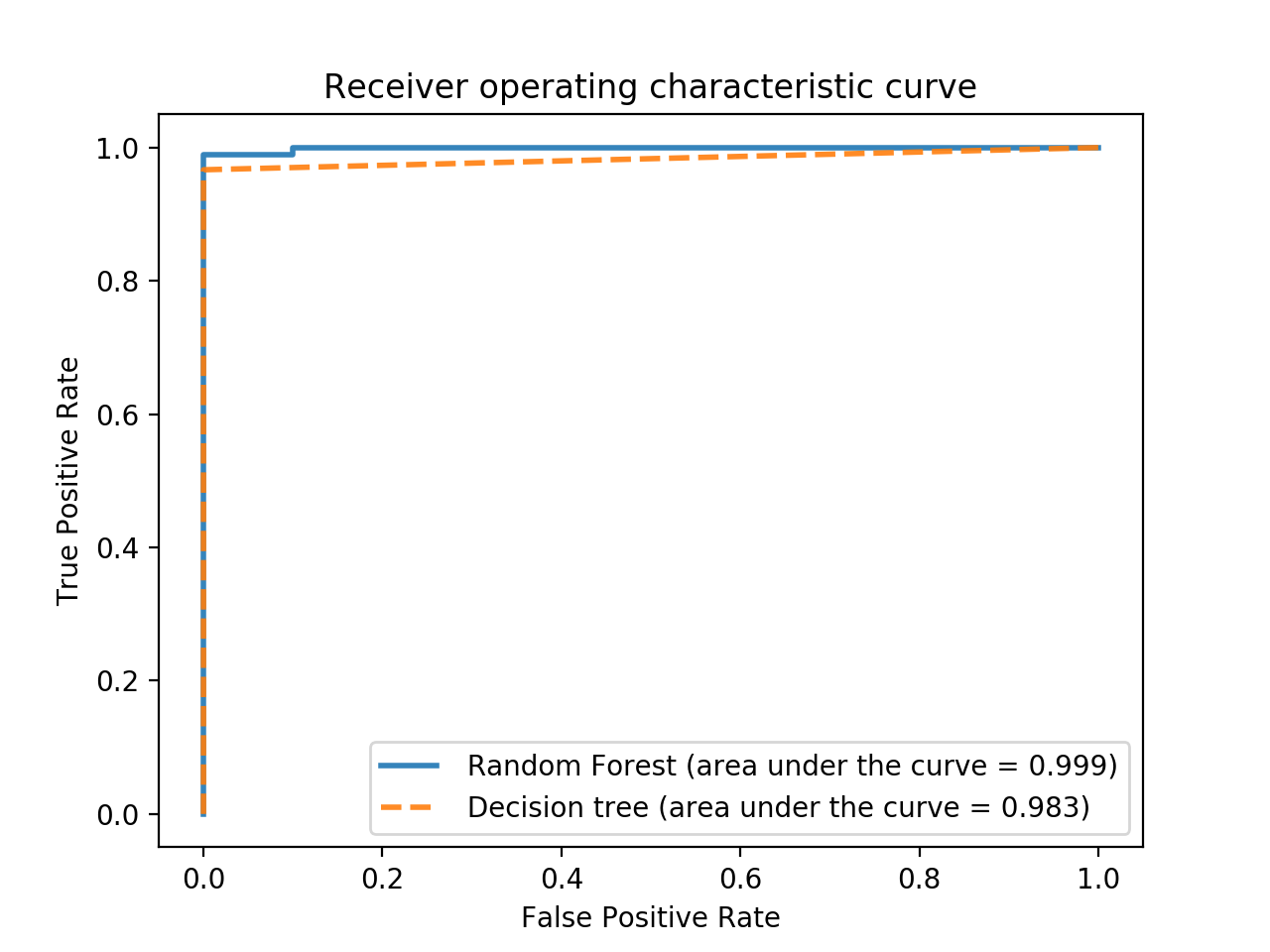}
\caption{An ROC curve for test sample for both the Decision Tree and Random Forest Classifier for comparison. It is clearly seen that the latter provides a better accuracy in the classification results. In addition, the respective areas under the curve are shown in the legend for both the methods.\label{fig:roc}}
\end{figure*}

\begin{figure*}
	\centering
	\makebox[\textwidth]{
	\includegraphics[width =0.5\textwidth]{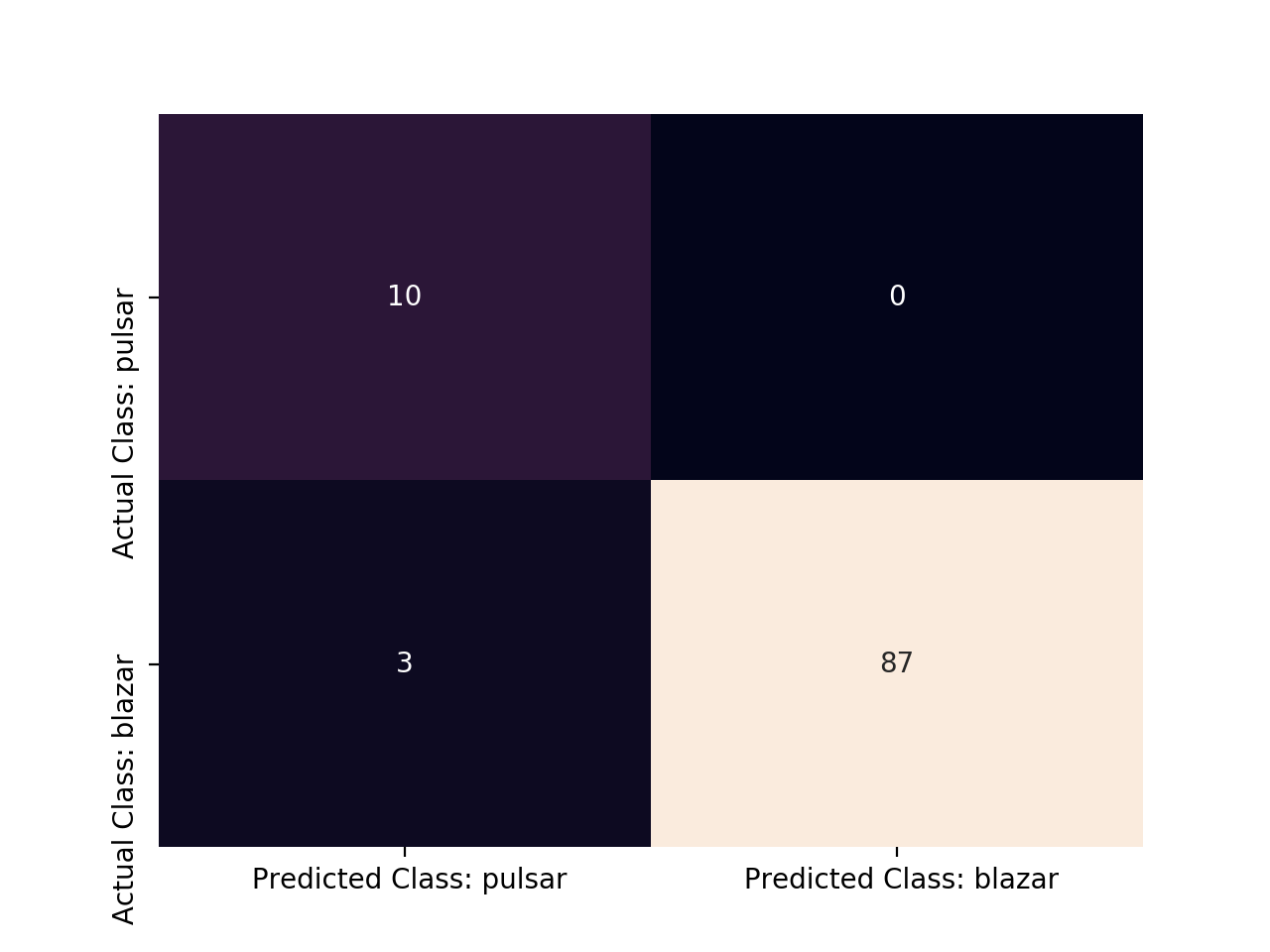}\hfill
	\includegraphics[width = 0.5\textwidth]{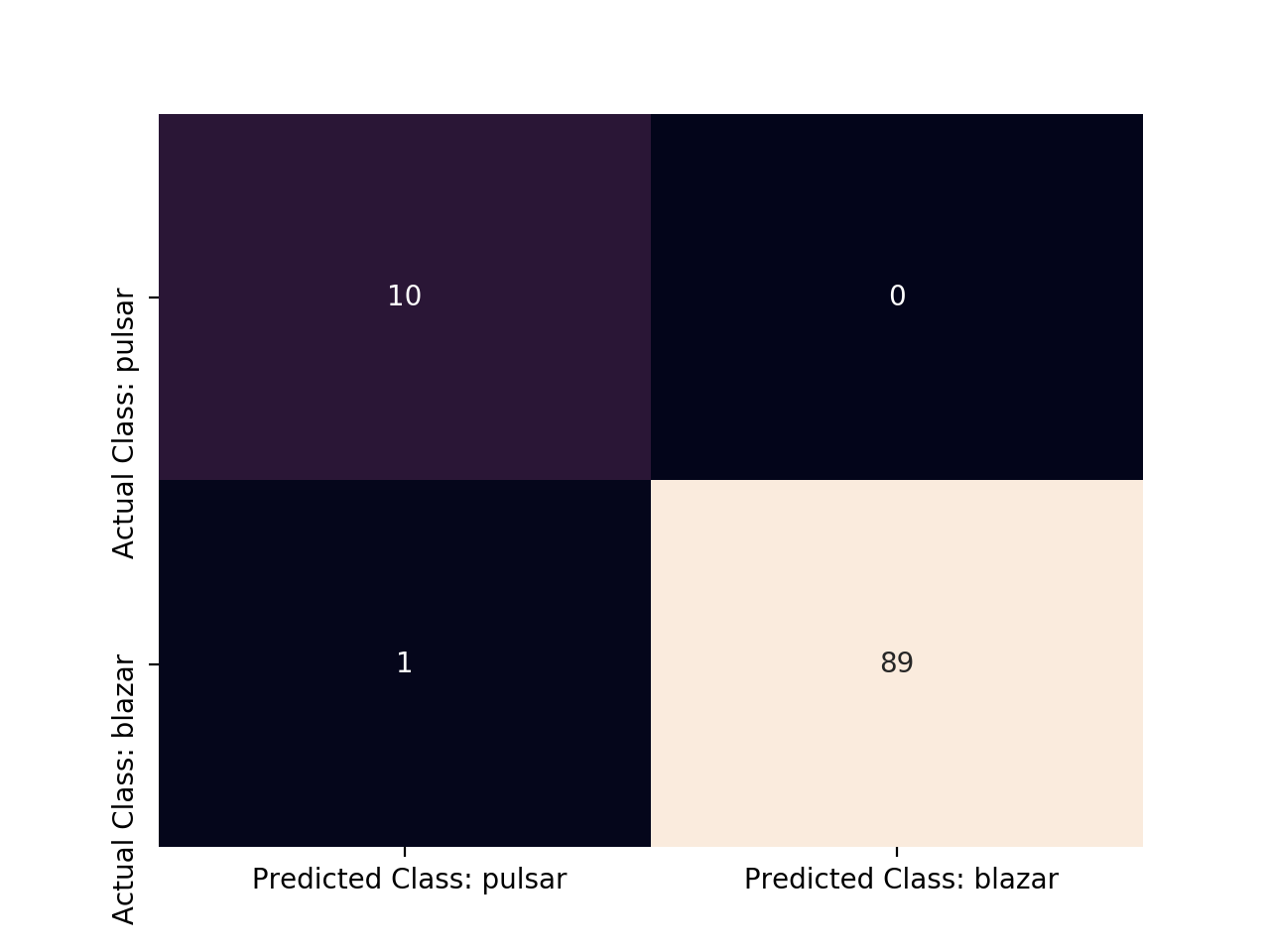}}
		\caption{(a) Confusion matrix for test sample (100 sources; 90 blazars and 10 pulsars)for the decision tree classifier. As seen from the figure, the decision tree predicted all pulsars correctly, but three blazars were wrongly predicted as pulsars. The accuracy of this method was 97\%. (b) Confusion matrix for test sample for the Random Forest Classifier. As seen from the figure, both the blazars and pulsars were correctly predicted by this method for 99 sources out of 100. Only one blazar was wrongly predicted as a pulsar in this case, yielding an accuracy score of 99\%. \label{fig:conf_matrix}}
\end{figure*}

The trained classifiers from both methods were finally applied to the sample of 217 X-ray sources, which yielded 39 candidate pulsars and 178 candidate blazars according to the single iteration of a decision tree classifier. The random forest classifier, which was based on 1000 decision tree iterations, predicted 13 likely pulsar candidates and 173 likely blazar candidates, assuming the sources with blazar probabilities $\geq$ 90\% are blazars and the ones with blazar probabilities $\leq$ 10\% are pulsars. The sources with P$_{bzr}$ $\geq$ 99\% and $\leq$ 1\% are termed as blazar candidates and pulsar candidates, respectively.  See Table~\ref{tab:ml} for details. The rest of the sources exhibiting ''ambiguous'' classification (31 in number), with blazar probabilities between 10\% and 90\%, are listed in Table~\ref{tab:amb}. The probability results from the RF classifier as well as our classification based on these probabilities are provided in each table. A receiver operating characteristic (ROC) curve, which displays the true positive rate vs false positive rate at various thresholds was constructed for both the methods. An ROC curve following a path more close to the left-hand border (small False Positive Rate) and then the top border (True Positive Rate 1) would be represent an ideal method with 100\% accuracy. In our case, RF yields slightly better accuracy than the DT method. See Fig.~\ref{fig:roc} for a comparison. In addition confusion matrices were generated for both the methods. A confusion matrix provides a visualization of the performance of the underlying algorithm provided true classification is known for that dataset. See Fig.~\ref{fig:conf_matrix} for details.  We emphasize that the results form a random classifier which is the iteration of 1000 decision trees are more robust as compared to a single decision tree run for classification as can be seen from both ROCs as well as confusion matrices.\\
Since the release of the 3FGL catalog, various independent studies led to identification/characterization of some of these sources. In particular, various optical spectroscopic campaigns, such as  \citet{Sandrinelli2013,Massaro2016a,Crespo2016b,Pena-Herazo2017,Paiano2017a,Paiano2017b} and \citep{Paiano2018} associated 56 
of these sources with QSOs, BL Lacs and Seyfert type 2 galaxies. Several others were identified as pulsars or pulsar candidates through multi-wavelength techniques and machine learning methods, respectively. In addition, the 4FGL catalog \citep{Collaboration2019} has been released this year which has identified 42 sources from our sample; 7 BL Lacs (BL Lacertae Objects, 7 FSRQs (Flat Spectrum Radio Quasars), 6 pulsars and 22 BCUs (Blazar Candidate of Uncertain Type) among these unassociated sources. See column 5 of Table~\ref{tab:ml} and \ref{tab:amb} and for details of these findings. Please note that all the possible classifications resulting from our machine learning algorithms with associated probabilities $\geq$ 99\% or $\leq$ 1\% are consistent with the results from independent studies. However, we note that two Fermi sources, 3FGL J0158.6+0102 and 3FGL J1322.3+0839 have been identified as a BL Lacs with an optical spectroscopic survey by \citet{Paiano2017b}, whereas they are identified as FSRQs in the 4FGL catalog. In addition, one source, 3FGL J1227.9-4834, which is listed as an ambiguous source according to our classification mechanism, has been previously identified as a low-mass X-ray binary (LXMB).\\
\subsection{Miscellaneous}
Out of the total 217 sources, we found that 3 sources, 3FGL J0748.8-2208, 3FGL J1624.1-4700, and 3FGL J1801.5-7825 have possible X-ray counterparts that are in positional coincidence with known stars within their respective uncertainties provided by the \Swift-XRT. In the case of 3FGL J1801.5-7825, this star is a K III subgiant, HD162298, which belongs to the category of FK Com stars. These stars are known as X-ray emitters due to their rapid rotation and strong magnetic fields. For 3FGL J1624.1-4700, the positionally coincident star is a rotationaly variable star, CD-46 10711. These stars could be associated with the coincident X-ray source, and the source of gamma-rays (e.g. as companions in low mass X-ray binary systems), or the positional overlap of the possibly associated sources could simply be a coincidence. The spectral type of the star, TYC 5993-3722-1, coincident with the \Swift XRT position for 3FGL J0748.8-2208 is unknown. It is possible that this star could be companion in a X-ray binary system or in a coincidental positional overlap with a background blazar. Please see Table~\ref{tab:ml}.
\section{Discussion and Conclusions}\label{sec:conclusion}
The main objective of this paper is to attempt to classify potential X-ray counterpart sources for the unassociated sample in the 3FGL catalog, which constitutes about one-third of the total source list. A complete classification of these mysterious gamma-ray sources is required for complete understanding of the high-energy universe. In this work, we utilize gamma-ray data in conjunction with X-ray data to classify these sources as either blazars or pulsars, since these two classes dominate the the known sources in the \Fermi catalogs. As already discussed, blazars can often be distinguished from pulsars based on just the gamma-ray and X-ray properties. We conduct a robust analysis by comparing a set of distinguishing parameters simultaneously using machine learning techniques. This analysis yields $\sim$ 79\% blazars and 6\% pulsars along with 14\% ambiguous sources. This is roughly consistent with the known gamma-ray source population in the \Fermi catalogs, and it has yielded several classifications of potentially new X-ray source associations with previously unassociated gamma-ray sources. From Table~\ref{tab:ml}, it can be seen that 134 of the likely X-ray/gamma-ray counterpart sources are identified as $\ge$99\% likely to be a blazar, with 75 of these not previously discovered or  classified. Similarly, out of the 7 pulsars based on $P_{bzr} \leq$ 1\%, 4 are new candidates based on our algorithm and the other 3 are listed as pulsars in the 4FGL catalog.\\
It should be noted that this study does not take into account the presence of other source classes, such as supernova remnants, globular clusters, starburst galaxies, high mass X-ray binaries, etc. It is indeed possible that some of the unassociated sources are neither blazars nor pulsars, in particular the ones with blazar probabilities less than 90\% and greater than 10\%. See Table~\ref{tab:amb}. In order to further confirm the classifications for these objects, in future work, we will (i) add more X-ray parameters derived from the spectral analysis, and (ii) utilize the information from other multi-wavelength catalogs, e.g. Wide-field Infrared Survey point source catalog \cite{Cutri2013}, NVSS\citep{Condon1998}, SUMSS\citep{Mauch2003}, ATCA\citep{Petrov2013}, UVOT, along with the gamma-ray and X-ray properties. The multiwavelength studies for these objects will indeed confirm the nature of the underlying sources, which would fit them into either blazar or pulsar or ''other'' categories. The mysterious sources in the ''other'' category are excellent targets for more thorough investigations. 
\appendix
\section{Pulsar Analysis from Swift Archival Data}
Out of 59 pulsars used in our machine learning algorithms, 10 were obtained from Swift archival data. Their spectra were fitted with both powerlaw and powerlaw with exponential cutoff models using XSpec \texttt{version 12.10.0c}. The column densities for all the sources were calculated using the HEASARC column density calculator\footnote{https://heasarc.gsfc.nasa.gov/cgi-bin/Tools/w3nh/w3nh.pl} and were fixed during the fitting procedure. The results from the best fit models are provided in the table below.
\begin{table*}[h]
\centering
\tablewidth{\textwidth}
\begin{tabular}{|C|C|C|C|C|C|C|C|}
\hline
  \multicolumn{1}{|c|}{3FGL} &
  \multicolumn{1}{c}{Swift OBS ID} &
  \multicolumn{1}{c|}{N$_{H}$} &
  \multicolumn{1}{c|}{$\Gamma_{X}$} &
  \multicolumn{1}{c|}{$\beta$} &
  \multicolumn{1}{c|}{Flux\tablenotemark{a}} &
  \multicolumn{1}{c|}{$\chi^{2}$} &
  \multicolumn{1}{c|}{d.o.f.} \\
\hline
J0205.5+6448 & 00010028003 & 0.48 & 1.80\pm0.15 & \nodata & 0.21 & 9.35 & 10\\
  J0437.2-4713 & 00080960001 & 0.01 & 2.85\pm0.05 & \nodata & 0.15 & 54.87 & 42\\
   J0534.5+2201 & 00058970001 & 0.21 & 1.89\pm0.03 & \nodata & 641.41 & 303.54 & 171\\
 J1119.1-6127 & 00081966001 & 1.09 & 1.41\pm0.18 & \nodata & 2.14 & 10.26 & 9\\
 J1227.9-4854 & 00041135011 & 0.11 & 1.53\pm0.16 & \nodata & 0.28 & 2.48 & 7\\
  J1509.4-5850 & 00080517002 & 1.66 & 1.61\pm0.07 & \nodata & 3.12 & 65.90 & 55\\
   J1823.7-3019 & 00035341002 & 0.13 & 1.01\pm0.007 & \nodata & 21.32 & 1043.12 & 725\\
   J1824.6-2451 & 00032785004 & 0.19 & 0.008\pm0.14 & 3.55\pm0.65 & 2.42 & 107.14 & 97\\
   J1833.5-1033 & 00053600099 & 1.25 & 0.13\pm0.16 & 2.38\pm0.28 & 8.31 & 142.04 & 149\\
  J2032.2+4126 & 00093148014 & 1.19 & 1.84\pm0.23 & \nodata & 0.44 & 1.96 & 6\\
\hline\end{tabular}
\tablenotetext{a}{The flux range is 0.1-2.4 keV and units are 10$^{-11}$ ergs/cm$^{2}$/s}
\end{table*}
\begin{longrotatetable}
\begin{deluxetable*}{CCcCCCl}
\tablecaption{Classification with Machine Learning \label{tab:ml}}
\tabletypesize{\scriptsize}
\tablecolumns{7}
\tablehead{
\colhead{Swift Name} & \colhead{Fermi Name} &\colhead{Class} &
 \colhead{Random Forest} & \colhead{X-ray Flux$^{\dagger}$}&\colhead{Gamma-ray Flux$^{\dagger}$}& \colhead{Notes}\\
\colhead{SwF3}  &\colhead{3FGL} & 
\colhead{} & \colhead{Blazar Probability}&\colhead{(0.1-2.4) keV}&\colhead{(0.1-100) GeV}&  \colhead{Classification in literature}} 
\startdata
 J000132.8-415523&J0002.2-4152&blazar&0.995&23.75&13.11& \\
J000805.3+145019&J0008.3+1456&blazar&0.999&28.27&16.11& BLL \citep{Paiano2017b}, bcu \citep[4FGL, ][]{Collaboration2019}\\
J000922.4+503029&J0009.3+5030&blazar&1&4.17&159.34& \\
J003119.8+072450&J0031.3+0724&blazar&0.999&6.26&15.8& \\
J003159.9+093616&J0031.6+0938&likely blazar&0.944&4.13&6.79& NLSy1  \citep{Paiano2017b}\\
J004859.4+422349&J0049.0+4224&blazar&1&6.91&16.72& BLL  \citep{Paiano2017c}\\
J011619.9-615343&J0116.3-6153&blazar&0.999&2.86&22.06& \\
J012152.5-391545&J0121.8-3917&likely blazar&0.971&31.76&11.56& BLL  \citep{Pena-Herazo2017}\\
J013106.8+612035&J0131.2+6120&blazar&0.993&118.9&118.45& \\
J013255.1+593213&J0133.3+5930&likely blazar&0.97&12.21&14.38& \\
J013320.9-441310&J0133.0-4413&blazar&1&3.37&16.41& bll \citep[4FGL, ][]{Collaboration2019}\\
J013750.3+581411&J0137.8+5813&blazar&0.993&139.6&49.49& \\
J014347.5-584552&J0143.7-5845&likely blazar&0.977&168.9&62.87& BLL \citep{Landoni_2015}\\
J015624.4-242003&J0156.5-2423&blazar&1&11.19&11.82& BLL  \citep{Pena-Herazo2017}\\
J015852.4+010127&J0158.6+0102&blazar&0.991&1.39&7.75& BLL  \citep{Paiano2017b}, fsrq \citep[4FGL, ][]{Collaboration2019}\\
J020020.9-410934&J0200.3-4108&blazar&0.998&8.02&15.75& BLL  \citep{Pena-Herazo2017} \\
J021210.5+532140&J0212.1+5320&likely pulsar&0.017&10.25&83.78&  pulsar \citep{Li2016} \\
J022302.7+682158&J0223.3+6820&likely blazar&0.989&19.4&31.75& \\
J022613.7+093725&J0226.3+0941&likely blazar&0.98&1.23&24.65& fsrq \citep[4FGL, ][]{Collaboration2019}\\
J023854.1+255406&J0239.0+2555&blazar&0.998&15.6&11.28& BLL   \citep{Paiano2017c}\\
J025047.7+562935&J0250.6+5630&blazar&0.998&22.41&31.19& \\
J025111.4-183115&J0251.1-1829&likely blazar&0.967&5.94&13.77& BLL  \citep{Paiano2017b}\\
J025857.5+055243&J0258.9+0552&blazar&0.996&5.98&26.3& BLL  \citep{Paiano2017b}\\
J030514.8-160820&J0305.2-1607&blazar&0.997&20.63&16.6& BLL  \citep{Paiano2017c}\\
J031614.2-643731&J0316.2-6436&blazar&0.997&62.52&31.08& BLL \citep{Landoni_2015} \\
J033514.0-445945&J0335.3-4459&blazar&0.995&4.99&32.5& \\
J033829.2+130215&J0338.5+1303&likely blazar&0.964&26.12&53.87& BLL  \citep{Paiano2017c}\\
J034050.0-242259&J0340.4-2423&blazar&0.999&3&11.64& QSO   \citep{Pena-Herazo2017}, bcu \citep[4FGL, ][]{Collaboration2019}\\
J034819.8+603507&J0348.4+6039&blazar&0.999&101.7&17.85& \\
J035051.2-281633&J0351.0-2816&blazar&0.999&30.24&10.16& BLL  \citep{Pena-Herazo2017}\\
J035309.4+565430&J0352.9+5655&blazar&0.996&27.14&37.64& BLL  \citep{Crespo2016}\\
J035939.3+764628&J0359.7+7649&blazar&0.994&4.93&10.47& bcu \citep[4FGL, ][]{Collaboration2019}\\
J040946.5-035958&J0409.8-0358&likely pulsar&0.908&3.13&38.07& BLL  \citep{Paiano2017c}\\
J041433.2-084214&J0414.9-0840&blazar&0.997&2.12&9.44& BLL  \citep{Paiano2017b}\\
J042011.0-601505&J0420.4-6013&blazar&0.993&20.01&15.97& BLL  \citep{Pena-Herazo2017}\\
J042749.8-670435&J0427.9-6704&blazar&0.993&3.91&21.36& \\
J042958.7-305932&J0430.1-3103&blazar&0.999&7.64&9.56& \\
J043836.8-732920&J0437.7-7330&likely blazar&0.986&3.69&13.63& \\
J043949.6-190100&J0439.9-1859&likely blazar&0.985&2.43&26.89& \\
J044722.5-253937&J0447.1-2540&blazar&0.996&3.04&11.14& BLL  \citep{Pena-Herazo2017}, bcu \citep[4FGL, ][]{Collaboration2019}\\
J045149.6+572141&J0451.7+5722&blazar&0.99&4.45&13.8& \\
J050650.1+032400&J0506.9+0321&blazar&0.999&6.25&14.99& BLL  \citep{Paiano2017b}\\
J051641.4+101243&J0516.6+1012&blazar&1&3.95&15.39& \\
J052140.9+010256&J0521.7+0103&blazar&0.997&1.06&21.69& \\
J053357.3-375755&J0533.8-3754&likely blazar&0.962&4.24&14.03& fsrq \citep[4FGL, ][]{Collaboration2019} \\
J055940.6+304233&J0559.8+3042&blazar&0.997&3.2&24.64& \\
J064847.6+151623&J0648.8+1516&blazar&0.993&197.9&86.49& \\
J065845.2+063711&J0658.6+0636&blazar&0.995&5.72&20.27& \\
J070014.4+130425&J0700.2+1304&blazar&0.998&11.38&23.73& BLL  \citep{Crespo2016}\\
J070421.7-482645&J0704.3-4828&blazar&0.999&9.9&10.43& \\
J072547.5-054830&J0725.7-0550&blazar&0.997&24.51&22.69& \\
J074627.0-022552&J0746.4-0225&blazar&0.998&14.24&31.49& \\
J074724.8-492634&J0747.5-4927&blazar&0.999&12.47&17.03& BLL  \citep{Pena-Herazo2017}\\
J074903.8-221016\tablenotemark{a}&J0748.8-2208&blazar&0.999&7.16&18.25&  \\
J080215.8-094214&J0802.3-0941&blazar&0.997&7.67&25.41& \\
J081338.1-035717&J0813.5-0356&blazar&0.995&29.42&17.09& \\
J082628.2-640416&J0826.3-6400&blazar&0.995&163.9&13.78& BLL \citep{Pena-Herazo2017}\\
J082930.3+085820&J0829.3+0901&blazar&1&2.31&14.64& fsrq \citep[4FGL, ][]{Collaboration2019} \\
J084121.3-355505&J0841.3-3554&blazar&0.998&23.48&106.29& \\
J084831.8-694109&J0847.2-6936&blazar&0.996&13.47&10.77& \\
J092818.1-525700&J0928.3-5255&likely blazar&0.984&8.27&23.01& \\
J093754.5-143349&J0937.9-1435&blazar&1&3.27&17.61& BLL  \citep{Paiano2017c}\\
J095249.5+071330&J0952.8+0711&blazar&0.999&6.93&17.83& BLL  \citep{Paiano2017c}, bcu\citep[4FGL, ][]{Collaboration2019}\\
J102432.6-454429&J1024.4-4545&blazar&0.999&29.91&13.23& \\
J103332.4-503527&J1033.4-5035&blazar&0.997&17.95&46.65& \\
J103755.1-242546&J1038.0-2425&likely blazar&0.929&4.12&11.79& bcu \citep[4FGL, ][]{Collaboration2019}\\
J104031.7+061722&J1040.4+0615&blazar&1&3&52.07& \\
J104503.3-594102&J1045.1-5941&pulsar&0.006&62.56&535.09& \\
J104939.4+154839&J1049.7+1548&likely blazar&0.985&6.99&15.92& bll \citep[4FGL, ][]{Collaboration2019}\\
J110506.3-611602&J1105.2-6113&blazar&0.9&3.15&93.04& pulsar \citep[4FGL, ][]{Collaboration2019}\\
J111715.1-533815&J1117.2-5338&blazar&0.999&7.26&44.36& \\
J111957.0-264322&J1119.8-2647&blazar&0.998&4.08&16.46& \\
J111958.9-220457&J1119.9-2204&pulsar&0.009&0.83&73.95& \\
J112504.2-580540&J1125.1-5803&likely blazar&0.988&22.21&23.27& \\
J112624.8-500807&J1126.8-5001&likely blazar&0.989&11.56&18.34& \\
J113032.6-780107&J1130.7-7800&likely blazar&0.985&141.8&30.49& \\
J113209.3-473854&J1132.0-4736&blazar&0.995&55.61&19.5& \\
J114141.7-140755&J1141.6-1406&likely blazar&0.988&23.8&18.48& BLL   \citep{Ricci2015}, bll \citep[4FGL, ][]{Collaboration2019}\\
J114600.8-063851&J1146.1-0640&blazar&0.999&9.2&17.5& BLL  \citep{Paiano2017b}\\
J114912.0+280720&J1149.1+2815&blazar&0.993&1.88&9.01& \\
J115514.5-111125&J1155.3-1112&likely blazar&0.988&4.6&15.97& \\
J120055.1-143039&J1200.9-1432&likely blazar&0.987&7.9&14.25& bll \citep[4FGL, ][]{Collaboration2019}\\
J122014.4-245948&J1220.0-2502&blazar&0.996&27.69&12.96& \\
J122019.8-371414&J1220.1-3715&blazar&0.996&15.34&21.24& \\
J122127.4-062846&J1221.5-0632&blazar&0.993&3&30.99& QSO  \citep{Crespo2016b}\\
J122257.0+121439&J1223.2+1215&blazar&0.998&0.9&15.85& bcu \citep[4FGL, ][]{Collaboration2019}\\
J122336.8-303247&J1223.3-3028&blazar&0.999&25.72&13.94& \\
J122536.7-344724&J1225.4-3448&blazar&1&30.36&12.59& \\
J123140.3+482149&J1231.6+4825&blazar&0.995&2.87&10.31& fsrq \citep[4FGL, ][]{Collaboration2019}\\
J123204.2+165528&J1232.3+1701&blazar&0.996&2.55&17.76& bll \citep[4FGL, ][]{Collaboration2019}\\
J123235.9-372056&J1232.5-3720&blazar&0.999&4.6&20.22& \\
J123447.7-043254&J1234.7-0437&blazar&0.99&3.23&16.04& Sy2  \citep{Paiano2017b}\\
J123726.6-705140&J1236.6-7050&blazar&1&5.23&20.21& \\
J124021.3-714858&J1240.3-7149&blazar&0.99&147.6&42.95& \\
J124919.5-280834&J1249.1-2808&blazar&0.995&34.16&24.57& \\
J124919.7-054540&J1249.5-0546&blazar&0.999&3.89&11.48& bcu \citep[4FGL, ][]{Collaboration2019}\\
J125058.4-494444&J1251.0-4943&blazar&0.993&2.77&25.55& \\
J125606.1-591931&J1256.1-5919&blazar&0.999&3.44&32.48& \\
J125949.4-374857&J1259.8-3749&blazar&0.993&3.45&27.85& BLL  \citep{Ricci2015}\\
J130059.5-814810&J1259.3-8151&likely blazar&0.988&3.48&16.65& \\
J131140.3-623314&J1311.8-6230&blazar&0.994&1.46&90.04& \\
J131552.8-073304&J1315.7-0732&blazar&0.998&21.83&42.6& \\
J132210.3+084230&J1322.3+0839&blazar&0.998&4.66&15.73& BLL  \citep{Crespo2016}, fsrq \citep[4FGL, ][]{Collaboration2019}\\
J132939.6-610735&J1329.8-6109&likely pulsar&0.059&4.26&82.45& \\
J134042.0-041009&J1340.6-0408&blazar&1&9.22&21.47& BLL  \citep{Paiano2017c}, bll \citep[4FGL, ][]{Collaboration2019}\\
J134706.8-295843&J1346.9-2958&blazar&0.99&14.45&32.72& BLL  \citep{Ricci2015}\\
J135340.2-663958&J1353.5-6640&blazar&1&98.07&47.41& \\
J140514.7-611823&J1405.4-6119&likely pulsar&0.053&6.54&364.56& \\
J141133.3-072256&J1411.4-0724&blazar&0.997&4.55&15.79& BLL  \citep{Paiano2017c}\\
J141901.2+773229&J1418.9+7731&likely blazar&0.937&29.31&25.19& \\
J144544.5-593200&J1445.7-5925&blazar&0.996&23.37&57.41& \\
J151148.6-051348&J1511.8-0513&blazar&0.994&181.8&42.29& BLL  \citep{Paiano2017c}\\
J151150.9+662450&J1512.3+6622&blazar&0.997&17.77&8.45& \\
J151212.9-225507&J1512.2-2255&blazar&0.999&12.35&33.85& BLL  \citep{Pena-Herazo2017}, bcu \citep[4FGL, ][]{Collaboration2019}\\
J151256.6-564027&J1512.8-5639&blazar&0.998&9.7&54.01& bcu \citep[4FGL, ][]{Collaboration2019}\\
J151319.0-372015&J1513.3-3719&blazar&0.993&3.99&15.38& \\
J151649.8+263635&J1517.0+2637&blazar&0.999&2.52&8.19& \\
J152603.0-083146&J1525.8-0834&blazar&0.995&4.21&11.27& BLL  \citep{Paiano2017b}, bcu\citep[4FGL, ][]{Collaboration2019}\\
J152818.2-290257&J1528.1-2904&blazar&0.999&6.37&12.47& bcu \citep[4FGL, ][]{Collaboration2019}\\
J154150.1+141441&J1541.6+1414&blazar&0.999&3.38&16.37& BLL  \citep{Paiano2017b}\\
J154459.2-664148&J1545.0-6641&likely blazar&0.975&99.02&25.03& \\
J154946.4-304502&J1549.9-3044&blazar&0.997&14.11&20.16& \\
J154952.1-065909&J1549.7-0658&blazar&1&47.5&51.58& \\
J161543.0-444921&J1615.6-4450&likely blazar&0.985&8.98&26.6& \\
J162432.2-465756\tablenotemark{b}&J1624.1-4700&likely pulsar&0.049&35.43&23.69& \\
J165338.2-015837&J1653.6-0158&pulsar&0&1.29&128.17& pulsar \citep[4FGL, ][]{Collaboration2019}\\
J170409.6+123423&J1704.1+1234&blazar&0.994&24.13&18.82& BLL  \citep{Paiano2017c}\\
J170433.9-052841&J1704.4-0528&likely blazar&0.977&35.56&34.16& BLL  \citep{Paiano2017c}\\
J171107.0-432416&J1710.6-4317&blazar&0.997&13.67&38.93& \\
J172142.1-392205&J1721.8-3919&blazar&0.998&12.77&60.06& \\
J172858.2+604400&J1729.0+6049&blazar&0.995&3.82&8.46& \\
J173250.5+591234&J1732.7+5914&blazar&1&3.9&8.94& \\
J180106.8-782248\tablenotemark{c}&J1801.5-7825&blazar&0.999&4.17&14.21&  \\
J181720.4-303258&J1817.3-3033&blazar&0.993&15.26&18.63& \\
J182338.8-345413&J1823.6-3453&likely blazar&0.964&284.6&113.07& \\
J183539.5+135048&J1835.4+1349&blazar&0.992&3.13&14.56& bll \citep[4FGL, ][]{Collaboration2019}\\
J184230.1-584158&J1842.3-5841&blazar&1&105.9&32.46& \\
J184433.1-034627&J1844.3-0344&pulsar&0.005&1.21&197.44& pulsar \citep[4FGL, ][]{Collaboration2019}\\
J190843.2-012954&J1908.8-0130&likely  pulsar&0.058&2.76&55& \\
J192114.1+194004&J1921.6+1934&likely blazar&0.964&15.13&26.68& \\
J192242.1-745355&J1923.2-7452&blazar&1&37.95&26.49& BLL  \citep{Pena-Herazo2017}\\
J193320.2+072620&J1933.4+0727&blazar&0.99&44.32&30.17& \\
J193420.1+600138&J1934.2+6002&blazar&0.996&7.35&15.7& bcu \citep[4FGL, ][]{Collaboration2019}\\
J194247.5+103327&J1942.7+1033&likely blazar&0.919&90.96&148.22& \\
J194633.6-540235&J1946.4-5403&pulsar&0.005&1.77&46.91& pulsar \citep[4FGL, ][]{Collaboration2019}\\
J195149.7+690719&J1951.3+6909&likely blazar&0.978&4.06&5.34& \\
J195800.3+243804&J1958.1+2436&blazar&0.996&24.16&24.55& \\
J200505.5+700437&J2004.8+7003&blazar&1&48.6&38.69& \\
J200635.7+015222&J2006.6+0150&likely blazar&0.965&4.13&24.17& pulsar \citep[4FGL, ][]{Collaboration2019}\\
J201431.1+064851&J2014.5+0648&blazar&1&20.16&35.62& \\
J201525.3-143205&J2015.3-1431&blazar&1&5.04&16.18& BLL  \citep{Crespo2016b}\\
J202154.9+062914&J2021.9+0630&blazar&0.996&2.36&27.83& BLL  \citep{Crespo2016}, bcu \citep[4FGL, ][]{Collaboration2019}\\
J203027.9-143919&J2030.5-1439&blazar&0.997&4.9&13.81& \\
J203450.9-420038&J2034.6-4202&blazar&0.999&15.01&20.59& \\
J203556.9+490038&J2035.8+4902&blazar&0.999&9.18&32.78& \\
J203649.6-332829&J2036.6-3325&likely blazar&0.955&45.79&16.75& BLL  \citep{Crespo2016b}\\
J203935.8+123002&J2039.7+1237&blazar&0.998&2.77&9.54& \\
J204312.6+171019&J2043.2+1711&pulsar&0.004&1.54&149.36& \\
J204351.5+103408&J2044.0+1035&likely blazar&0.923&4.5&16.94& bcu \citep[4FGL, ][]{Collaboration2019}\\
J205357.9+690518&J2054.3+6907&likely blazar&0.985&1.08&18.17& \\
J205950.4+202905&J2059.9+2029&likely blazar&0.983&5.04&8.43& \\
J210940.0+043958&J2110.0+0442&blazar&0.995&8.98&16.64& \\
J211522.2+121802&J2115.2+1215&blazar&0.996&3.59&15.16& \\
J211754.9-324329&J2118.0-3241&blazar&1&5.2&11.72& \\
J212729.3-600102&J2127.5-6001&blazar&1&20.1&10.02& bcu \citep[4FGL, ][]{Collaboration2019}\\
J212945.1-042907&J2129.6-0427&likely pulsar&0.091&1.91&30.86& pulsar \citep[4FGL, ][]{Collaboration2019}\\
J213348.6+664704&J2133.8+6648&blazar&1&7.14&57.88& \\
J214247.5+195812&J2142.7+1957&blazar&1&12.8&10.23& \\
J215123.0+415635&J2151.6+4154&blazar&0.996&18.46&38.15& \\
J220941.7-045109&J2209.8-0450&likely blazar&0.926&3.04&15.14& BLL  \citep{Paiano2017b}\\
J221532.1+513529&J2215.6+5134&pulsar&0.002&1.41&73.41& \\
J222911.2+225456&J2229.1+2255&blazar&0.99&54.31&13.32& BLL  \citep{Paiano2017b}, bcu \citep[4FGL, ][]{Collaboration2019}\\
J224437.0+250344&J2244.6+2503&blazar&1&3.42&13.59& BLL  \citep{Paiano2017b}\\
J224710.1-000512&J2247.2-0004&blazar&0.99&0.72&26.93& BLL  \citep{Sandrinelli2013}\\
J225003.5-594520&J2249.3-5943&likely blazar&0.962&2.62&9.67& \\
J225032.7+174918&J2250.3+1747&blazar&0.991&1.98&15.86& BLL  \citep{Paiano2017b}, bcu \citep[4FGL, ][]{Collaboration2019}\\
J230012.4+405223&J2300.0+4053&likely blazar&0.984&18.22&19.72& \\
J230351.7+555618&J2303.7+5555&blazar&0.995&30.74&23.73& \\
J230848.5+542612&J2309.0+5428&blazar&0.998&5.03&14.52& \\
J232127.1+511118&J2321.3+5113&blazar&1&5.73&11.7& \\
J232137.1-161926&J2321.6-1619&blazar&0.993&26.84&11.78& BLL  \citep{Paiano2017b}\\
J232938.7+610112&J2329.8+6102&blazar&0.996&44.15&29.13& \\
J233626.4-842650&J2337.2-8425&blazar&0.997&6.73&14.16& BLL  \citep{Pena-Herazo2017}\\
J235115.9-760018&J2351.9-7601&blazar&0.997&7.98&17.73& BLL  \citep{Pena-Herazo2017}\\
J235825.0+382857&J2358.5+3827&blazar&1&20.47&18.5& Sy2  \citep{Paiano2017b}\\
J235836.8-180718&J2358.6-1809&blazar&1&23.48&18.97& BLL  \citep{Paiano2017b}\\
\enddata
\tablenotetext{a}{positionally coincident with a star, TYC 5993-3722-1}
\tablenotetext{b}{positionally coincident with a rotationally variable star, CD-46 10711 of type K1IV(e)}
\tablenotetext{c}{positionally coincident with a star, HD162298 of type K4III}
\tablenotetext{\vdag}{Flux in the units of 10$^{-13}$ erg/cm$^2$/s)}

\end{deluxetable*}
\end{longrotatetable}

\startlongtable
\begin{deluxetable*}{CCCl}
\tablecaption{Classification using Machine Learning : Ambiguous classifications \label{tab:amb}}
\tabletypesize{\scriptsize}
\tablecolumns{4}
\tablewidth{\columnwidth}
\tablehead{
\colhead{Swift Name} & \colhead{3FGL Name} &
 \colhead{Random Forest} & \colhead{Notes}\\
\colhead{SwF3}  &\colhead{3FGL} & 
 \colhead{Blazar Probability}& \colhead{Classification in literature}} 
\startdata
J052939.5+382321 &  J0529.2+3822 & 0.121 &\\
J082623.6-505743 &  J0826.3-5056  & 0.198 & \\
   J083843.4-282702 &  J0838.8-2829  & 0.116 & \\
   J085505.8-481518 &  J0855.4-4818 & 0.14 & \\
   J085755.9-483424 &  J0858.0-4834  & 0.176 & \\
   J093444.6+090356 &  J0935.2+0903  & 0.692 & \\
   J112042.3+071313 &  J1120.6+0713  & 0.124 & bcu \citep[4FGL,][]{Collaboration2019}\\
   J122758.7-485342 &  J1227.9-4854  & 0.417 & XSS J12270-4859 \citep{DeMartino2015}\\
   J125821.5+212352 &  J1258.4+2123  & 0.228 & \\
   J130832.0+034407 &  J1309.0+0347  & 0.59 & \\
   J141045.2+740505 &  J1410.9+7406  & 0.154 & \\
   J142035.9-243022 &  J1421.0-2431  & 0.348 & \\
   J154343.6-255608 &  J1544.1-2555  & 0.178 & \\
   J162607.8-242736 &  J1626.2-2428c   & 0.15 & \\
   J173508.3-292954 &  J1734.7-2930  & 0.255 & \\
   J175316.4-444822 &  J1753.6-4447  & 0.123 & \\
   J175359.6-292908 &  J1754.0-2930  & 0.106 & \\
   J180351.7+252607 &  J1804.1+2532  & 0.34 & \\
   J180425.0-085003 &  J1804.5-0850  & 0.874 & \\
   J181307.6-684713 &  J1813.6-6845  & 0.572 & \\
   J182914.0+272902 &  J1829.2+2731  & 0.131 & bcu \citep[4FGL,][]{Collaboration2019}\\
   J182915.5+323432 &  J1829.2+3229  & 0.145 & bcu \citep[4FGL,][]{Collaboration2019}\\
   J184833.8+323251 &  J1848.6+3232  & 0.73 & \\
   J185606.6-122148 &  J1856.1-1217  & 0.518 & \\
   J190444.5-070743 &  J1904.7-0708  & 0.77 & \\
   J201537.2+371119 &  J2015.6+3709  & 0.862 & FSRQ \citep[4FGL,][]{Collaboration2019}\\
   J204806.3-312012 &  J2047.9-3119  & 0.781 & bcu \citep[4FGL,][]{Collaboration2019}\\
   J212601.5+583148 &  J2125.8+5832  & 0.222 & \\
   J214429.5-563850 &  J2144.6-5640  & 0.614 & BLL \citep{Pena-Herazo2017}\\
   J215046.5-174956 &  J2150.5-1754  & 0.504 & BLL \citep{Paiano2017b}, bcu \citep[4FGL,][]{Collaboration2019}\\
   J225045.6+330515 &  J2250.6+3308  & 0.151 & \\
\enddata
\end{deluxetable*}


\clearpage

\software{scikit-python \citep[version 0.20.3,][]{scikit-learn},Topcat \citep[version 4.6-3,][]{Taylor2005}}
\acknowledgements
The authors would like to gratefully acknowledge the support provided 
by NASA research grants 80NSSC17K0752 and 80NSSC18K1730. This research has made use of the ZBLLAC spectroscopic library. http://www.oapd.inaf.it/zbllac. The astronomical tool to compare databases, Topcat \citep{Taylor2005} was employed in this work. We would like to thank Dr. Eric Feigelson at Pennsylvania State University for the help and feedback in the implementation of the machine learning methods.
\bibliography{my_bibliography}
\end{document}